\documentclass[journal=jacsat,manuscript=article,hyperref=true]{achemso}

\usepackage{amssymb, amsmath, subcaption, caption, mathrsfs, url}
\usepackage{lipsum}
\usepackage[utf8]{inputenc}
\usepackage[T1]{fontenc}
\usepackage{graphicx}
\usepackage{multirow}
\usepackage[export]{adjustbox}
\usepackage{xcolor}
\usepackage{comment}
\usepackage{siunitx}

\usepackage[version=3]{mhchem}

\author{Yasir Mahmood}
\affiliation{Material Measurement Laboratory, National Institute of Standards and Technology (NIST),
Gaithersburg, MD 20899, USA}
\alsoaffiliation{Department of Chemistry and Biochemistry, University of Maryland, College Park, USA}
\email{yasir.mahmood@nist.gov}

\author{Daniel Wines}
\affiliation{Material Measurement Laboratory, National Institute of Standards and Technology (NIST),
Gaithersburg, MD 20899, USA}

\author{Viviana Dovale-Farelo}
\affiliation{Material Measurement Laboratory, National Institute of Standards and Technology (NIST),
Gaithersburg, MD 20899, USA}
\alsoaffiliation{Department of Chemistry and Biochemistry, University of Maryland, College Park, USA}

\author{Kamal Choudhary}
\affiliation{Material Measurement Laboratory, National Institute of Standards and Technology (NIST),
Gaithersburg, MD 20899, USA}
\alsoaffiliation{Department of Materials Science and Engineering, Whiting School of Engineering, The Johns Hopkins University, Baltimore, MD 21218, USA}
\alsoaffiliation{Department of Electrical and Computer Engineering, Whiting School of Engineering, The Johns Hopkins University, Baltimore, MD 21218, USA}

\author{Lucas M Hale}
\affiliation{Material Measurement Laboratory, National Institute of Standards and Technology (NIST),
Gaithersburg, MD 20899, USA}
\email{lucas.hale@nist.gov}

\title{Strain-Induced Interfacial Bond Evolution and \textcolor{msc1}{Facet-dependent Strain Response of Electronic Properties} in Oxidized Silicon Nanowires}

\keywords{Nanowires, Straintronics, ReaxFF}

\definecolor{msc1}{HTML}{000000}
\definecolor{msc2}{HTML}{000000}
\definecolor{msc}{HTML}{000000}

\begin{document}

\begin{abstract}
Silicon nanowires (Si NWs) have emerged as promising candidates for next-generation nanoelectronic devices due to their tunable electronic properties, and compatibility with aggressive scaling requirements. The integration of strain further enhances their performance by modulating carrier mobility, band structure, and transport behavior without altering material chemistry. Within the NW, the silicon-silicon dioxide (Si-SiO\textsubscript{2}) interface serves as a critical region where strain redistribution and suboxide bonding collectively influence the mechanical and electronic response. Here, we combine reactive force field molecular dynamics and first-principles calculations to investigate how oxidation-induced interfacial structure and crystallographic facet orientation govern strain-dependent bond evolution and \textcolor{msc1}{facet-dependent strain response of electronic properties} in oxidized Si NWs. Simulated dry thermal oxidation generates chemically realistic Si-SiO\textsubscript{2} interfaces, enabling direct characterization of bond order redistribution and stress gradient development across the interface under tensile strain. Interfacial Si-Si bonds exhibit elongation and reduced bond order associated with partial oxidation states, leading to localized stress concentration and preferential bond rupture. Complementary density functional theory calculations on facet-specific models reveal that orientation-dependent interfacial bonding modulates \textcolor{msc1}{facet-dependent strain response of electronic properties}. Together, these results establish a direct coupling between interfacial suboxide chemistry, mechanical deformation, and electronic structure, providing atomistic insight into strain-modulated behavior at oxidized silicon interfaces in faceted NWs.
\end{abstract}
\section{Introduction}
Continuous scaling of semiconductor devices has driven the transition from planar Field-Effect Transistors (FETs) to three-dimensional architectures such as FinFETs and, more recently, Gate-All-Around (GAA) FETs, which provide improved electrostatic control in nanoscale channels~\cite{ferain2011multigate}. In GAA architectures, silicon nanowires (Si NWs) serve as the channel material, where the gate fully surrounds the conduction path, enhancing electrostatic confinement and scalability~\cite{ferry2008nanowires}. At these reduced dimensions, the high surface-to-volume ratio renders material properties increasingly sensitive to atomic-scale structural features, particularly at the silicon-silicon dioxide (Si-SiO\textsubscript{2}) interface formed through oxidation~\cite{pan2008semiconductor}.

Beyond geometrical manipulation, strain provides a powerful means of modulating the electronic structure of Si through controlled lattice deformation~\cite{tsutsui2019strain}. By perturbing bond lengths and coordination environments, strain can modulate carrier mobility, band structure, and transport properties~\cite{niquet2012effects, greil2012tuning, shiri2008strain, kleimann1998stress}. In Si NWs, where interfacial effects are amplified, the response to strain is expected to depend sensitively on the atomic-scale bonding at the Si-SiO\textsubscript{2} boundary. \textcolor{msc1}{Although numerous theoretical and experimental studies have examined strain-induced electronic modulation in Si NWs~\cite{mohapatra2020strain, ghadimi2018elastic, niquet2012effects, bartmann2024electronic}, comparatively little attention has been directed toward understanding how interfacial suboxide bonding configurations evolve under externally applied mechanical strain and how these local bonding changes promote bond weakening, interfacial stress localization, and facet-dependent strain response of electronic properties}.

The Si-SiO\textsubscript{2} interface represents an atomically abrupt transition between crystalline Si and amorphous SiO\textsubscript{2}, which introduces residual stresses~\cite{lin1972local}, strained bonding configurations~\cite{jaccodine1966measurement}, and defect states~\cite{mehes2017defects}, all of which strongly influence both electronic and mechanical responses. Despite extensive investigations of oxidation chemistry and interface defects, quantitative understanding of how interfacial suboxide bonding and stress distributions evolve under applied strain, particularly at the nanoscale, remains limited. \textcolor{msc1}{Furthermore, the influence of crystallographic facet orientation on the electronic properties under strain has seldom received systematic attention despite its potential importance for directional charge transport and band-edge modulation in Si NW devices~\cite{badawy2024electronic, rurali2010colloquium}.} This gap limits a fundamental understanding of coupling between strain, local bonding characteristics and electronic structure at oxidized Si interfaces. \textcolor{msc1}{Si-SiO\textsubscript{2} interfaces may arise either through native oxidation under ambient exposure or through intentional thermal oxidation, and these two oxidation routes can produce distinct interfacial structures and properties. These differences can influence oxide morphology, suboxide bonding, residual stress, defect populations, and consequently the mechanical and electronic response of the interface~\cite{philipp1982optical}. In the present work, we focus specifically on Si-SiO\textsubscript{2} interfaces generated through simulated dry thermal oxidation.}

In this work, we combine large-scale reactive force field molecular dynamics~\cite{Aktulga12} (ReaxFF MD) with density functional theory (DFT) to resolve how oxidation-induced interfacial structure governs both mechanical and electronic responses of Si NWs. We demonstrate that under uniaxial loading, bond weakening and stress localization occur preferentially at the Si-SiO\textsubscript{2} interface, leading to consistent fracture initiation at this boundary. This mechanism remains robust across variations in oxide thickness and core size. To connect these atomistic insights with electronic functionality, we further perform first-principles calculations on facet-specific Si-SiO\textsubscript{2} interfaces, revealing strain-dependent \textcolor{msc1}{electronic properties}. Together, these results establish a direct coupling between interfacial suboxide chemistry, mechanical deformation, and electronic structure in oxidized Si NWs, providing guidance for the design of robust and strain-engineered NW based devices.
\section{Methods}
\subsection{Oxidation and Straining}
In this study, the Si-SiO\textsubscript{2} interface was constructed to closely resemble the structural and chemical characteristics in experimentally realized systems. Other computational studies have typically generated Si-SiO\textsubscript{2} interfaces by first creating amorphous SiO\textsubscript{2} (a-SiO\textsubscript{2}) via the melt-quench method~\cite{kluge1987amorphous}, followed by matching the a-SiO\textsubscript{2} layer with a crystalline Si (c-Si) substrate in conjunction with energy minimization~\cite{fischer2006structure, korkin2006computational, mehes2017defects, tit1999electronic, van2007atomistic, pasquarello1995si} or by sequentially inserting oxygen atoms into Si-Si bonds to mimic gradual oxidation~\cite{watanabe2004sio2}. \textcolor{msc1}{While these approaches are computationally efficient and can yield structurally reasonable oxide layers, the imposed construction procedure and/or insufficient relaxation may introduce artificial interfacial mismatch or local bonding artifacts. Ganster et al. showed that oxygen-insertion-based methods are sensitive to the insertion protocol, which can strongly influence the resulting oxide morphology and interfacial bonding environment~\cite{ganster2010molecular}. In the present work, we instead use a simulated dry thermal oxidation procedure in which O\textsubscript{2} molecules dynamically interact with the exposed Si surface and, after formation of an oxide layer, oxygen transport and reaction proceed through the growing oxide toward the Si-SiO\textsubscript{2} interface. This approach allows oxygen incorporation, Si-O bond formation, and the crystalline-to-amorphous transition to develop dynamically during oxidation.}
\par
\textcolor{msc1}{Low index faceting is common in Si NWs, but the physical origin of the exposed facets depends on the fabrication route. In bottom-up NW synthesis, faceting arises from the intrinsic energetic preference of the system to minimize surface energy by forming boundaries along low-energy crystallographic planes~\cite{zhang2005structures}. For instance, oxide-assisted growth typically favors elongation along the axial [110] and [112] directions~\cite{wang1998sio}, whereas NWs synthesized via the metal-catalyzed vapor-liquid-solid (VLS) method predominantly exhibit axial growth along the [111] direction~\cite{morales1998laser}. Using chemical vapor deposition (CVD) with gold nanoclusters as catalysts and silane (SiH\textsubscript{4}) as the vapor-phase reactant, Wu et al. reported a strong correlation between growth direction and wire diameter~\cite{wu2004controlled}. In the 3~nm - 10~nm range, 95~\% of Si NWs grew along the $\langle$110$\rangle$ direction; in the 10~nm - 20~nm range, 61~\% exhibited $\langle$112$\rangle$ orientation; and in the 20~nm - 30~nm range, 64~\% favored the $\langle$111$\rangle$ direction. However, in top-down FinFET  and GAA compatible Si NW channels, the axial orientation, cross-sectional shape, and exposed facets are governed primarily by wafer orientation, lithographic patterning direction, anisotropic or isotropic etching, and sacrificial oxidation rather than by growth kinetics. Previous studies have shown that top-down SiNW GAA devices may exhibit triangular, pentagonal, circular, rectangular, or other process-defined cross sections~\cite{pott2008fabrication, chen2009top, de2014top, barraud2019top}, and that NW orientation, cross-sectional geometry, and surface/facet orientation can significantly affect carrier transport~\cite{fujihara2015impacts,liang2007structure}. Based on these considerations, the [110] oriented NW used in this work was selected as a representative low index faceted Si geometry for subsequent oxidation simulations, rather than as a direct reproduction of either a specific bottom-up growth morphology or a single top-down device-fabrication process. The model exposes \{111\} and \{100\} surface families, enabling a controlled comparison of facet surfaces and facet junctions during oxidation, tensile loading, and electronic-structure analysis.}
\par
\begin{figure}[t!]
    \begin{center}
        \begin{tabular}{c}
            \includegraphics[width = 0.99\textwidth]{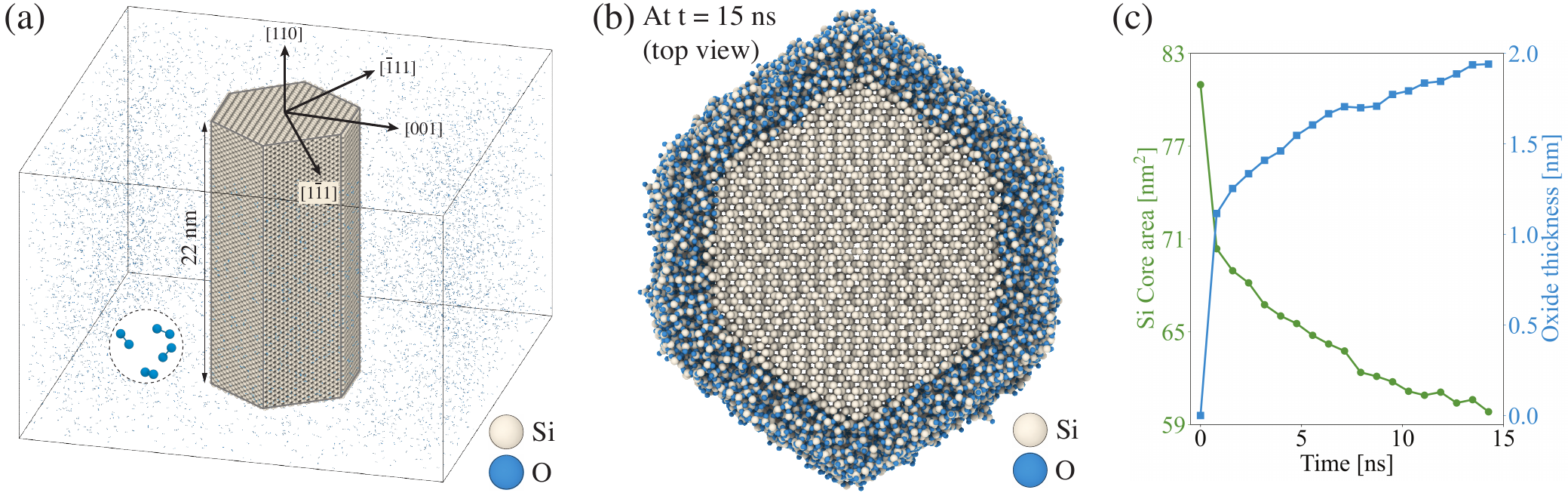}
        \end{tabular}
    \end{center}
    \caption{\label{fig:fig1} (a) Faceted Silicon Nanowire (Si NW) exposed to an oxygen environment (b) The top view of the same NW after 15~ns of oxidation, showing the consumption of the Si core and the formation of an amorphous oxide shell. (c) Temporal evolution of the Si core cross-sectional area (left axis) and the oxide shell thickness (right axis). The inverse correlation reflects the consumption of Si during oxidation and provides the basis for quantifying interface location and oxide thickness in subsequent analyses.}
\end{figure}

The MD simulations were performed using the Large-scale Atomic/Molecular Massively Parallel Simulator (LAMMPS) package~\cite{LAMMPS}. \textcolor{msc1}{Prior to oxidation, the pristine faceted Si NW, approximately 22 nm in length, was heated from 300 K to 1000 K over 1 ns and then equilibrated at 1000 K for 3 ns under an isothermal-isobaric (NPT) ensemble before introducing O\textsubscript{2} molecules. During this pre-oxidation equilibration, the barostat was applied only along the periodic NW axis at a target axial pressure of 0. This allowed the axial cell length to relax and reduced artificial axial stress associated with construction of the initial NW geometry. The transverse box dimensions were kept fixed, while the free NW surfaces were allowed to relax atomistically. The equilibrated wire was then exposed to an O\textsubscript{2} environment, where the molecular density was selected empirically to promote sustained oxidation kinetics. At 1000 K, the corresponding average atomic velocity was approximately 0.012 \AA/femtosecond. While a [110]-oriented NW can, in principle, expose multiple low-index surface configurations~\cite{zhang2005structures}, the present geometry exhibits six surface facets belonging to the \{111\} and \{100\} crystallographic plane families. This faceting is consistent with experimental reports for Si NWs of comparable orientation and diameter~\cite{wu2004controlled, ma2003small}. Fig.~\ref{fig:fig1}(a) shows the initial idealized faceted NW geometry before equilibration and oxidation.} \textcolor{msc1}{Because the free NW surfaces and facet junctions were allowed to relax during pre-oxidation equilibration and subsequently react dynamically with oxygen during oxidation, the oxidized structure used for tensile testing does not preserve perfectly sharp crystalline corners, but instead develops a rounded and disordered oxide/interface morphology, as shown in Fig.~\ref{fig:fig1}(b).}

The system is maintained at 1000 K to enhance oxidation dynamics and is evolved in a canonical ensemble (NVT) with periodic boundary conditions in the x, y, and z directions. The ReaxFF~\cite{Aktulga12} interatomic potential employed in this study is adopted from Noaki et al.~\cite{noaki2023development}. This potential has been benchmarked against key structural descriptors (bond lengths, angle distributions) and mechanical responses for a-SiO\textsubscript{2}. \textcolor{msc1}{A summary of these results, along with comparisons to other interatomic potentials and experimental observations, is presented in Table~\ref{tab:amorphousSiO2comparison}. Overall, the potential provides reasonable agreement with reported amorphous SiO\textsubscript{2} structural parameters, particularly for Si-O, Si-Si, and O-O distances and the O-Si-O angle. However, the average Si-O-Si angle is underestimated relative to experimental values. This underprediction is also reported in the original development and validation study of this ReaxFF parameter set, where the potential was optimized to capture a broader set of Si/O/H and oxidation relevant properties rather than the Si-O-Si angle alone.}
\par
The oxidation process proceeds in two stages. In the initial stage, O\textsubscript{2} molecules impinging on the exposed Si facets undergo thermally activated dissociation upon interaction with reactive surface sites. The resulting atomic oxygen preferentially inserts into surface Si-Si bonds, forming stable Si-O bonds and locally converting the crystalline Si into amorphous SiO\textsubscript{2}. Consequently, this stage is reaction-limited, with oxidation kinetics governed by the frequency of O\textsubscript{2} dissociation and direct oxygen incorporation.
In the subsequent stage, once a continuous SiO\textsubscript{2} layer of finite thickness forms, oxidation proceeds via oxygen diffusion through the amorphous oxide network towards the underlying Si-SiO\textsubscript{2} interface. Here, oxygen atoms must migrate through the silica, often involving bond switching and network rearrangement, before incorporating into Si atoms at the interface. This diffusion-limited regime slows the oxidation front propagation compared to the initial direct-reaction stage, consistent with experimental observations~\cite{watanabe1998kinetics} and Deal-Grove type kinetics in thermal oxidation~\cite{deal1965general}. In our simulations, this two-stage process results in the growth of an oxide shell that reaches 2 nm within 15 ns, as shown in Fig.~\ref{fig:fig1}(b). The corresponding temporal evolution of oxide growth and consumption of Si core is presented in Fig.~\ref{fig:fig1}(c), which highlights the rapid initial increase in oxide thickness followed by a marked reduction in rate due to the transition in oxidation mechanism.
\par
To investigate the mechanical response of oxidized Si NWs under uniaxial loading, we use the wire with 2~nm oxide shell thickness for our analysis.  \textcolor{msc1}{To eliminate thermal history effects from oxidation at 1000~K, the oxidized NW was linearly cooled from 1000~K to 300~K over 1~ns under NPT ensemble, allowing axial relaxation during cooling. The structure was then equilibrated again under NPT at 300~K for 10~ns to ensure structural relaxation and removal of residual stress associated with high temperature oxidation, thermal contraction, and oxide growth. The free NW surfaces were allowed to relax atomistically during the process.} The equilibrated wire was subsequently subjected to axial tension under displacement-controlled boundary conditions at 300~K. We investigated the mechanical response across a wide range of strain rates and evaluated their effect on stress-strain behavior (see Fig. S1 in the Supporting Information). Based on these results, a strain rate of $10^8~s^{-1}$ was selected for subsequent simulations. Although this strain rate remains several orders of magnitude higher than experimental conditions due to computational constraints, it is sufficiently slow to allow atomic rearrangement during loading and has been shown to capture elastic responses consistent with other computational studies~\cite{chowdhury2016molecular, muralidharan2007molecular}. \textcolor{msc1}{The axial strain reported in the MD tensile simulations is the global engineering strain, defined relative to the relaxed axial cell length of the oxidized NW before tensile loading.}
\begin{table}[t!]
\centering
\caption{\label{tab:amorphousSiO2comparison} Comparison of geometric parameters in amorphous SiO\textsubscript{2} from experimental data, previous MD studies, and the current MD simulation. All distances are in \AA~and angles in degrees.}
\begin{tabular}{lccc}
\hline\hline
Geometric Parameter & Experimental & Other MD Studies & Current MD \\
\hline
Si-O Bond Length (\AA)   & 1.62\cite{mozzi1969structure}     & 1.62\cite{vashishta1990interaction} & \textcolor{msc1}{1.60} \\
Si-Si Distance (\AA)     & 3.08\cite{konnert1973computation} & 3.00\cite{van2003reaxffsio}         & 3.02 \\
O-O Distance (\AA)       & 2.63\cite{wyckoff1966crystal}     & 2.59\cite{vollmayr1996cooling}      & 2.61 \\
Si-O-Si Angle ($^\circ$) & 153\cite{da1975refinement}        & 152\cite{vollmayr1996cooling}       & 144  \\
O-Si-O Angle ($^\circ$)  & 109.3\cite{mozzi1969structure}    & 108.3\cite{vollmayr1996cooling}     & 107  \\
\hline\hline
\end{tabular}
\end{table}
\subsection{Density Functional Theory Calculations}
To investigate the \textcolor{msc1}{strain response of electronic properties} in these NWs, we make the assumption that the electronic properties of planar interfaces can be representative of facet NW geometry. \textcolor{msc1}{We used Si-SiO\textsubscript{2} interfacial structures generated from the same simulated dry thermal oxidation procedure employed in the ReaxFF MD simulations. Representative periodic planar supercells were constructed for the relevant facet orientations and then oxidized using the same ReaxFF based dry thermal oxidation workflow. This procedure preserves the oxidation generated Si-SiO\textsubscript{2} bonding motifs and local interfacial disorder associated with thermal oxidation while reducing the system size to make first-principles calculations computationally feasible. The resulting DFT models therefore represent facet-resolved oxidized Si-SiO\textsubscript{2} interfaces generated through the same MD oxidation framework, rather than independently constructed ideal Si-SiO\textsubscript{2} slabs.} Specifically, we studied the electronic properties of planar facets \textcolor{msc1}{(001), (111), and (110)} by performing density functional theory (DFT)~\cite{PhysRev.136.B864,PhysRev.140.A1133} calculations with the Vienna {\it Ab initio} Simulation Package (VASP) using the projector augmented wave (PAW) method~\cite{PhysRevB.54.11169,PhysRevB.59.1758}. \textcolor{msc1}{The (001) and (111) interfaces represent the \{100\} and \{111\} surface family exposed by the MD NW geometry, while the (110) interface is added as an additional low-index Si-SiO\textsubscript{2} interface relevant to other possible process-defined SiNW geometries~\cite{zhang2005structures}.} We utilized the OptB88vdW~\cite{klimevs2011van} exchange correlation functional, which is a Generalized Gradient Approximation (GGA) that includes van der Waals (vdW) interactions. \textcolor{msc2}{It is important to note that GGA-level exchange correlation functionals systematically underestimate the band gap of semiconductors and insulators (i.e., Si and SiO$_2$), but important qualitative trends can be drawn from GGA-level electronic structure calculations. Due to this, we expect the relative (qualitative) trends in band gap across different facets to be valid despite this underestimation in magnitude.} The self-consistent convergence criteria for energy was set to 10$^{-6}$ eV. These calculations were performed with a kinetic energy cutoff of 600 eV and a \textcolor{msc1}{k-point grid of 3 x 3 x 1 for the (001) facet, 4 x 3 x 1 for the (111) facet, and 5 x 3 x 1 for the (110) facet.} \textcolor{msc1}{Using the same MD dry oxidation workflow, representative smaller periodic planar interface supercells were constructed for DFT calculations.} \textcolor{msc1}{The simulation cell for the (001) facet contained 366 atoms, the cell for the (111) facet contained 301 atoms, and the cell for the (110) facet contained 234 atoms.} In order to alleviate the well-known phenomena of dangling bonds in a-SiO$_2$ creating defect states within the band gap (resulting in Si-SiO$_2$ becoming a conductor), we passivated the dangling Si bonds with H. After fully relaxing the H-passivated facets with DFT, strain was applied in the $b$-direction [110], to mimic the strain state of the NW. This was done by increasing the lattice constant $b$  of the simulation cell and allowing the atoms to relax within the cell. \textcolor{msc1}{Static applied strain from -4\% to +4\%, in increments of 1\%, was applied to each facet resolved DFT interface model. The strain was defined as engineering strain relative to the relaxed axial cell dimension of the corresponding DFT interface model. At each applied strain, the atomic positions were relaxed before the electronic-structure calculations were performed.} \textcolor{msc1}{To approximate the electronic conductivity, we performed calculations using semi-classical Boltzmann transport theory with the BoltzTraP2 code \cite{MADSEN2018140}. BoltzTraP2 uses the constant relaxation time approximation and rigid band approach to calculate transport properties \cite{MADSEN2018140}. In this approximation, scattering physics and electron-phonon coupling are not explicitly taken into account. A full solution of the Boltzmann transport equation would require phonon dispersions and electron-phonon coupling constants, which are computationally intractable for the systems in this work. Our approximated conductivity is effectively a band-structure-weighted transport coefficient, $\sigma/\tau_0$, which cannot be converted to an absolute conductivity without computing $\tau_0$. Due to this, only qualitative comparisons can be made for electronic conductivity.} \textcolor{msc2}{The Python Materials Genomics (pymatgen) \cite{ONG2013314} package was used to extract the trace-averaged $\sigma/\tau_0$, where $\tau_0 = 10^{-14}~s$ (the default value in pymatgen).} Additional supporting DFT calculations were performed to quantify the change in bond length and bond order at the interface and support the findings of ReaxFF simulations. These are discussed in the SI, where additional methodology details are given.

\section{Results and Discussion}
The Si-SiO\textsubscript{2} interface was found to strongly influence the mechanical integrity of oxidized Si NW, owing to the abrupt transition between crystalline and amorphous bonding environments. To probe its role in mechanical failure, we focus on rupture initiation and propagation under tensile loading using multiple independent replica simulations. While the strain levels leading to rupture exceed those encountered in device operation, analyzing the fracture regime provides a sensitive means of identifying the weakest atomic environments and reveals the interfacial mechanisms that govern mechanical reliability. \textcolor{msc1}{Additionally, the rupture strain values reported here should be interpreted as idealized MD global engineering fracture strains, not as homogeneous elastic strain states throughout the crystalline Si lattice. Prior in situ TEM/MD comparison studies have shown that MD can reproduce the brittle fracture mechanism of Si NWs, including surface- or defect-mediated crack initiation and rapid crack propagation, while often overpredicting absolute fracture stresses and strains because simulated systems are smaller, cleaner, and contain fewer pre-existing defects than experimental samples~\cite{tang2012mechanical, delrio2016near}. In addition, MD tensile simulations are necessarily performed at strain rates much higher than those used experimentally, which can further increase the predicted fracture stress and fracture strain. Comparable large global fracture or yield strains have also been reported in displacement-controlled atomistic tensile simulations of Si NWs and oxidized Si systems~\cite{chowdhury2016molecular, sun2019effect, yang2009shape}. Therefore, the present values are used to identify the atomistic fracture mechanism and weakest interfacial bonding environments, rather than to claim quantitative agreement with experimental fracture strain values.}

\begin{figure}[t!]
    \begin{center}
        \begin{tabular}{c}
            \includegraphics[width = 0.99\textwidth]{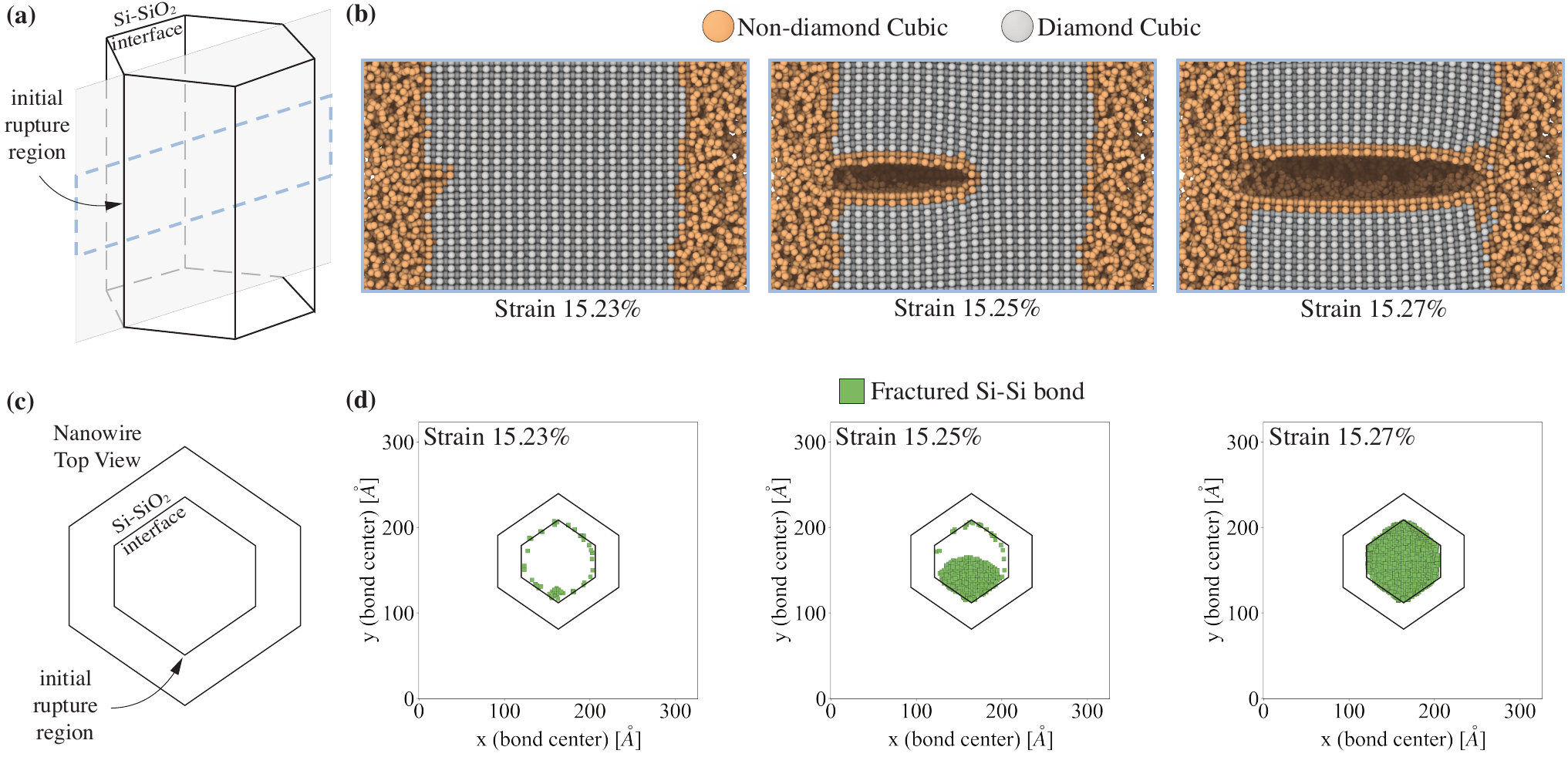}
        \end{tabular}
    \end{center}
    \caption{\label{fig:nw_rupture}Progression of Si NW core rupture under uniaxial strain. (a) Schematic of the NW slice used for side-view visualization. (b) Atomistic side-view snapshots showing fracture evolution at strain levels of 15.23~\%, 15.25~\%, and 15.27~\%. \textcolor{msc1}{Atoms are colored according to their local coordination, distinguishing diamond-cubic (gray) from non-diamond-cubic (orange).} (c) Schematic of the NW top view used for bond-breaking analysis. (d) Corresponding top-view plots of fractured Si-Si bonds, identified relative to the last intact configuration at 15~\% strain. The sequence highlights rupture initiation at the Si-SiO\textsubscript{2} interface, propagation through the Si core, and final rupture across the axial plane.}
\end{figure}

Across all 24 independent replicas, the rupture consistently initiated at the Si-SiO\textsubscript{2} interface, either along an individual facet or at the junction between two facets. The corresponding stress-strain responses of the replicas are provided in Fig. S2 of the Supporting Information. The first fracture events appeared as a localized group of Si-Si bond failures at the interface, marking the onset of rupture before propagating into the Si core. Fig.~\ref{fig:nw_rupture}(a) shows the schematic of the cross-sectional slice used for visualization, while Fig.~\ref{fig:nw_rupture}(b) presents atomistic snapshots of the sliced NW at increasing strain. Complementary top-view schematics are shown in Fig.~\ref{fig:nw_rupture}(c), with the corresponding fractured-bond maps in Fig.~\ref{fig:nw_rupture}(d), where broken bonds are identified relative to the intact reference configuration at 15~\% strain. Together, these panels illustrate rupture initiation at 15.23~\% strain, propagation at 15.25~\%, and complete Si core rupture by 15.27~\% strain. \textcolor{msc1}{Additionally, as shown in Fig.~\ref{fig:nw_rupture}(b), after rupture of the crystalline Si core, portions of the surrounding amorphous oxide shell remain partially connected across the fracture region and continue to carry a reduced tensile load during further deformation, thereby explaining the residual post-rupture stress observed in Fig.~S1.}

\par
\begin{figure}[t!]
    \begin{center}
        \begin{tabular}{c}
            \includegraphics[width = 0.99\textwidth]{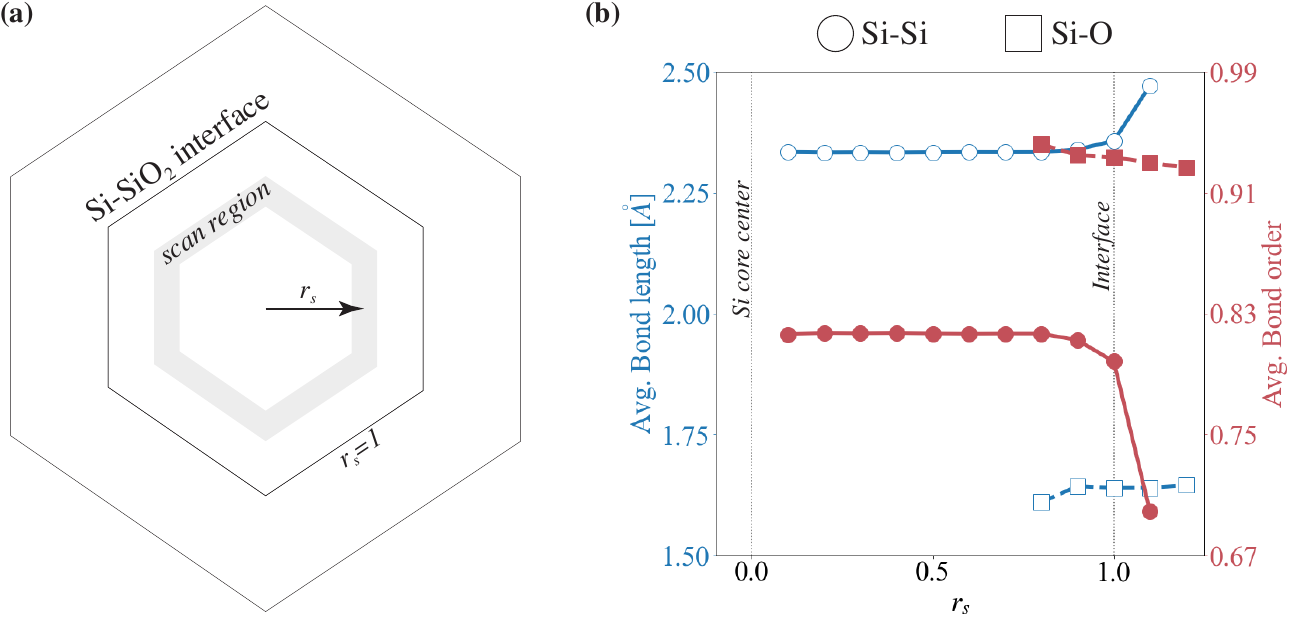}
        \end{tabular}
    \end{center}
    \caption{\label{fig:bo_bl_annulus}(a) Top view of the NW cross-section illustrating the definition of the radial sampling annulus (grey region). The dimensionless parameter $r_s$ is used to normalize the distance of the scan region from the Si-SiO\textsubscript{2} interface where $r_s<1$ corresponds to Si core, $r_s=1$ marks the interface, and $r_s>1$ represents the SiO\textsubscript{2} shell. The shaded annulus denotes the region sampled for averaging bond lengths and bond orders at a given $r_s$. (b) The variation of average bond length and bond order for the Si-Si and Si-O bonds within the scan region, plotted as a function of the radial scaling parameter $r_s$.}
\end{figure}

To elucidate how the oxidation modifies the local bonding environment, we analyze the variation in Si-Si and Si-O bond lengths and bond orders as a function of radial position across the Si-SiO\textsubscript{2} interface. In ReaxFF, the bond order between atoms $i$ and $j$ is defined as~\cite{van2003reaxffsio}:
\begin{equation}
\label{eq:bo}
BO_{ij} = exp\left[p_{bo1}\left(\frac{r_{ij}}{r_o^\sigma}\right)^{p_{bo2}}\right] + 
exp\left[p_{bo3}\left(\frac{r_{ij}}{r_o^\pi}\right)^{p_{bo4}}\right] + exp\left[p_{bo5}\left(\frac{r_{ij}}{r_o^{\pi\pi}}\right)^{p_{bo6}}\right]
\end{equation}
where $r_{ij}$ is the interatomic distance, $r_o^\sigma$, $r_o^\pi$, $r_o^{\pi\pi}$ are reference bond lengths for $\sigma$, $\pi$ and double-$\pi$ interactions respectively, and $p_{bo}$ are empirical fitting parameters. Physically, higher bond orders correspond to stronger bonds, whereas oxidation-induced charge transfer and overcoordination tend to reduce bond order, particularly in partially oxidized environments. Tracking these variations thus provides a quantitative measure of how interfacial oxidation redistributes local valence and modifies the effective bond strength. Full details of the ReaxFF bond-order expression are available in Ref.~\cite{van2003reaxffsio}.
For computing average bond characteristics, we divided the NW cross-section into concentric sampling annuli of width 8~\AA, fitted to the interface location. Given the NW circumcircle diameter 14~nm (with oxide shell), this annulus thickness provides adequate radial resolution while ensuring enough atom counts for statistical averaging. The interface location is defined by fitting a convex hull over oxygen atoms with maximum penetration depth.  The radial location of each annulus was normalized by the interface position to obtain a dimensionless radial scaling parameter $r_s$, where $r_s=1$ identifies the Si-SiO\textsubscript{2} interface, $r_s<1$ corresponds to the Si core and, $r_s>1$ denotes the oxide shell. Fig.~\ref{fig:bo_bl_annulus}(a) illustrates the sampling scheme, while Fig.~\ref{fig:bo_bl_annulus}(b) presents the corresponding radial variations of average Si-Si and Si-O bond lengths and bond orders.
\par
The radial trends reveal that Si-Si bonds near the interface undergo both bond elongation and reduction in bond order relative to those in the crystalline core. This indicates that interfacial Si-Si bonds are simultaneously strained and electronically weakened. In ReaxFF, bond order depends not only on interatomic distance but also on coordination environment and valence redistribution (see Eqn.~\ref{eq:bo}). At the interface, oxygen incorporation frequently produces partially oxidized Si species (Si\textsuperscript{1+}, Si\textsuperscript{2+}, Si\textsuperscript{3+}), which diverts valence toward Si-O bonding. As a result, nearby Si-Si bonds exhibit reduced covalent strength and elongate from an average of 2.33~\AA~in the core to approximately 2.47~\AA~at the interface. In contrast, Si-O bonds remain shorter ($\approx$~1.62~\AA) and exhibit higher bond orders, consistent with their stronger covalency. A slight decrease in Si-O bond order from the interface into the oxide shell is also observed, which we attribute to increasing structural disorder and variation in local coordination within amorphous SiO\textsubscript{2}. The combined reduction in Si-Si bond order and accumulation of partial suboxides thus identify the Si-SiO\textsubscript{2} interface as a mechanically vulnerable zone.
\par
To isolate the local electronic effect of oxygen incorporation on Si-Si bonding, we analyzed a simplified crystalline Si model at 0~K using DFT, comprising a 64 atom supercell with one to three interstitial O atoms. For each configuration (Si, Si\textsuperscript{1+}, Si\textsuperscript{2+}, Si\textsuperscript{3+}), we computed Si-Si and Si-O bond lengths and bond orders using the Chargemol code (DDEC6 analysis) \cite{chargemol}. The detailed methodology, including setting up the supercell and bond order evaluation is described in the SI. \textcolor{msc1}{Table~\ref{table_chargemol} summarizes the normalized average bond lengths and bond orders (relative to Si-Si bonds in pristine Si) for back Si-Si bonds obtained using DDEC6.} Introducing oxygen systematically redistributes valence from Si-Si to Si-O bonds, in line with prior first-principles studies of oxygen complexes in silicon \cite{needels1991oxygen}. The Si-O bonds remain short ($\approx$1.60~\AA) and strongly covalent across all oxidized configurations, whereas Si-Si bonds in the vicinity of oxygen exhibit reduced bond order relative to bulk-like Si, even when their geometric distance decreases, consistent with theoretical predictions that interstitial oxygen weakens neighboring Si-Si bonds \cite{plans1987theoretical}.

\textcolor{msc1}{For first nearest neighbor Si-Si bonds, the normalized bond orders are 0.93, 0.94, and 0.97 for Si\textsuperscript{1+}, Si\textsuperscript{2+}, and Si\textsuperscript{3+} environments, respectively. Although these values increase slightly with oxidation state, they remain below the pristine Si value. This non-monotonic behavior reflects a competition between two effects: oxidation-induced valence redistribution, which tends to reduce Si-Si bond order, and local nearest neighbor Si-Si bond contraction, which increases orbital overlap and partially compensates for charge depletion~\cite{saito1988stable}. While the second and third NN Si-Si bond orders also remain below the pristine Si value, they show a clearer decreasing trend with increasing oxidation state, decreasing from 0.976 to 0.953 for 2NN and from 0.953 to 0.933 for 3NN.}

This local electronic weakening couples to mesoscale strain around the Si-SiO\textsubscript{2} interface. As oxygen content increases, the average first-nearest-neighbor Si-Si distance near O decreases from 2.37~\AA\ in pristine Si to 2.22~\AA\ in the Si\textsuperscript{3+} system. This contraction indicates local densification of the partially oxidized shell, in agreement with experimental observations of compressive stress formation during Si NW oxidation~\cite{shir2006oxidation}. The local compression is accommodated by a progressive expansion of next-nearest-neighbor Si-Si distances beyond the bulk value, with a concomitant monotonic reduction in their normalized bond order (down to $\sim$0.93 for third neighbors in Si\textsuperscript{3+}). In other words, the Si atoms closest to oxygen bear both electronic depletion (lower Si-Si bond order) and geometric frustration (compressed first neighbors and stretched outer shells). Taken together with the ReaxFF trends across the NW cross-section, these DFT calculations support a unified picture in which partial oxidation creates a shell of Si\textsuperscript{1+}/Si\textsuperscript{2+}/Si\textsuperscript{3+} species that preferentially strengthen Si-O bonds at the expense of nearby Si-Si bonds, while inducing strain that further elongates and weakens more distant Si-Si linkages \cite{mehes2017defects}. This combined electronic and mechanical degradation around the Si-SiO\textsubscript{2} interface explains why rupture consistently nucleates at the interface region and identifies partially oxidized Si environments as intrinsically weak links in the load-bearing network. 
\begin{table}[t!]
  \centering
  \caption{\textcolor{msc1}{Average Si-Si and Si-O distances and DDEC6 Si-Si bond orders (BO) around oxidized Si atoms for different oxidation states.} Here, 1NN, 2NN, and 3NN denote first, second, and third nearest neighbors, respectively; both distances and bond orders are normalized with respect to the pristine Si system (bulk Si: $d=1.000$, BO = 1.000)}
  \label{table_chargemol}
  \begin{tabular}{lccccccc}
    \hline
    System 
    & $d_{Si-Si}^{1NN}$ 
    & $d_{Si-Si}^{2NN}$ 
    & $d_{Si-Si}^{3NN}$ 
    & $d_{Si-O}$
    & ${BO}_{Si-Si}^{1NN}$ 
    & ${BO}_{Si-Si}^{2NN}$ 
    & ${BO}_{Si-Si}^{3NN}$ \\
    \hline
    Si                     & 1.000 & 1.000 & 1.000 & --    & 1.000 & 1.000 & 1.000 \\
    Si\textsuperscript{1+} & 0.989 & 0.998 & 1.004 & 0.679 & 0.933 & 0.976 & 0.953 \\
    Si\textsuperscript{2+} & 0.971 & 0.998 & 1.005 & 0.679 & 0.938 & 0.974 & 0.948 \\
    Si\textsuperscript{3+} & 0.937 & 0.998 & 1.008 & 0.674 & 0.967 & 0.953 & 0.933 \\
    \hline
  \end{tabular}
\end{table}
\par
\begin{figure}[t!]
    \begin{center}
        \begin{tabular}{c}
            \includegraphics[width = 0.99\textwidth]{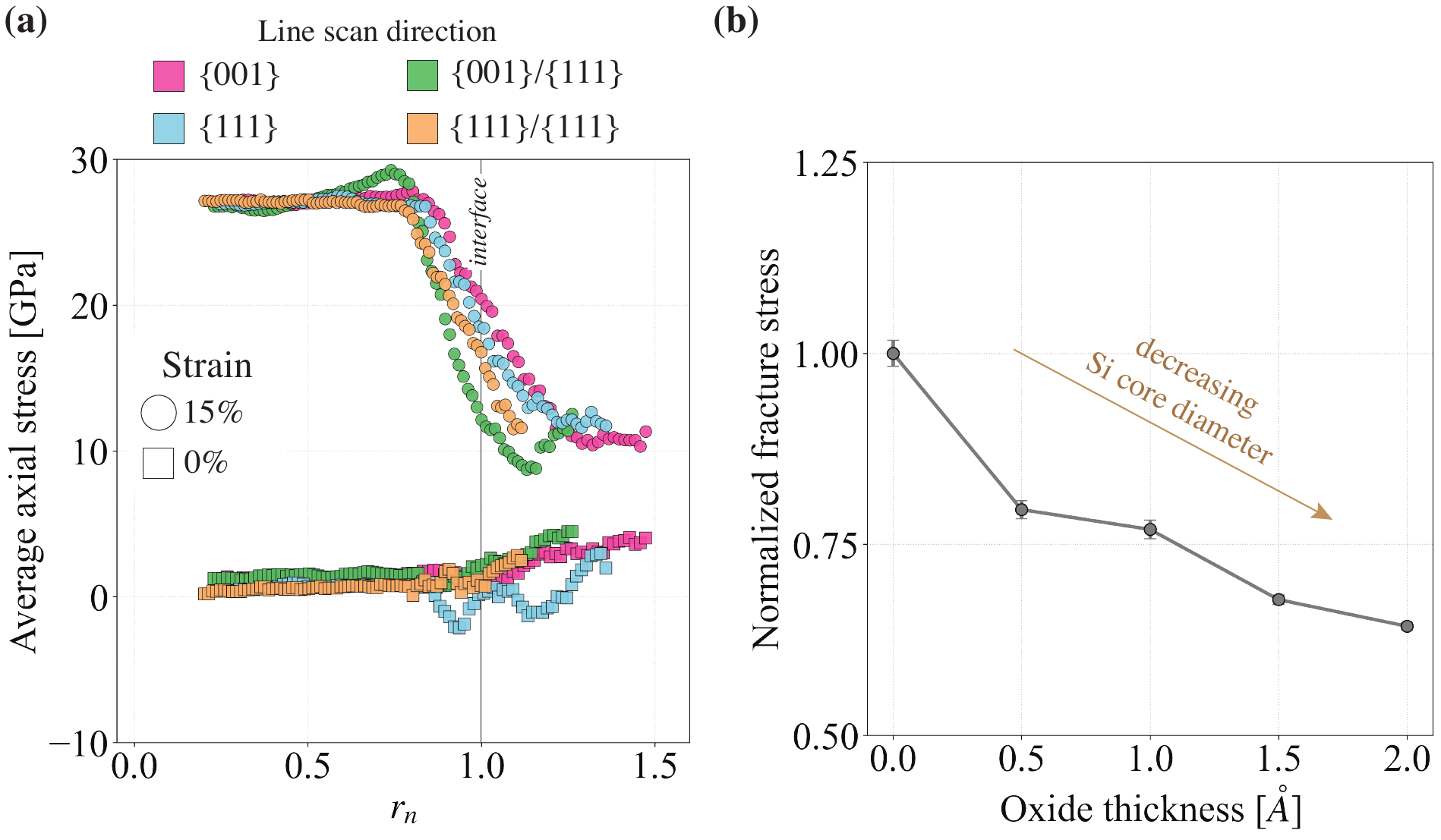}
        \end{tabular}
    \end{center}
    \caption{\label{fig:stress_LS} (a) Axial stress ($\sigma_{zz}$) line scans across the Si-SiO\textsubscript{2} interface at 0~\% strain and immediately prior to rupture (15~\% strain). $r_n$ is a dimensionless parameter. Results are shown for the four symmetry-unique directions along: \{001\}, \{001\}/\{111\} junctions, \{111\}, and \{111\}/\{111\} junctions. At 0~\% strain, all scans exhibit small oscillations in stress at the interface, which relax back to baseline values within the oxide shell. Just prior to rupture, however, a pronounced localized peak emerges along the \{111\}/\{111\} junction, coinciding with the site where fracture initiates (failure strain = 15.1~\%). (b) Normalized fracture stress of Si NW as a function of oxide shell thickness during progressive oxidation.}
\end{figure}

Next, we analyze the mechanical state of the Si-SiO\textsubscript{2} interface by performing directional line scans along radial vectors either normal to individual facets or passing through the junctions between adjacent facets (see Fig.~\ref{fig:fig1}(a) for reference). These scans resolve how axial stress ($\sigma_{zz}$) goes from the crystalline Si core through the interfacial region and into the oxide shell. At zero applied strain, the average virial stress is nearly uniform across the cross-section, though small oscillations are visible at the interface that relax back within the oxide. As tensile loading is applied, a gradient develops: stresses in the Si core rise more steeply than in the shell, with the oxide sustaining only about one-third of the load carried by the core, consistent with the relative stiffness (bulk moduli) of SiO\textsubscript{2} and Si. Representative results for facet-directed stress scans are shown in Fig.~\ref{fig:stress_LS}.
For each line scan, stresses were averaged over atoms within a sliding square sampling window of 16 x 16~\AA, advanced radially from the NW center toward the shell. The window size ensures sufficient atom counts while preserving radial resolution, and successive overlap provides smoothing of local fluctuations. The radial coordinate was normalized by the center-to-interface distance along the particular direction, such that the normalized radius $r_n=1$ marks the Si-SiO\textsubscript{2} interface, $r_n<1$ the crystalline core, and $r_n>1$ the oxide shell. This normalization allows direct comparison across the four symmetry-unique directions: (i) two normal to \{001\}, (ii) four normal to \{111\}, (iii) four along \{001\}/\{111\} junctions, and (iv) two along \{111\}/\{111\} junctions.

\textcolor{msc1}{As shown in Fig.~\ref{fig:stress_LS}(a), for the representative replica analyzed here, at strains just prior to fracture (15~\%), a localized stress concentration emerges at the Si-SiO\textsubscript{2} interface along the line scan direction passing through the subsequent rupture-initiation region. This small but distinct peak coincides with the site where rupture subsequently initiates. Across the independent replicas, the specific crystallographic location of rupture varies depending on the local oxidation topology and interfacial disorder. However, the common feature across the replicas is that rupture consistently initiates at the Si-SiO\textsubscript{2} interface where a local stress concentration develops. No comparable interfacial peaks are observed along other scan directions or cross-sections for the representative replica shown in Fig.~\ref{fig:stress_LS}(a), indicating that fracture is spatially localized rather than uniformly distributed.} \textcolor{msc1}{To further assess the stress state near rupture, we examined the instantaneous diagonal components of the virial pressure tensor for the same representative trajectory. The transverse components remain close to zero throughout the pre-rupture tensile regime, while the axial component follows the same trend as the axial stress–strain response. After core rupture, transient fluctuations appear in transverse direction reflecting local stress redistribution, oxide-shell deformation, and atomic rearrangement during fracture. These results confirm that the pre-rupture response is dominated by axial tensile loading.}

The fracture mechanism can therefore be attributed to localized interfacial stress concentrations that trigger rupture of the Si core. Within the range of oxide thicknesses tested (0~nm - 2~nm), this mechanism remains unchanged; the shell thickness primarily alters the overall stress level sustained by the NW rather than the fracture initiation site. Larger-scale simulations or slower strain rates may broaden the distribution of fracture strains by sampling a wider variety of interfacial defect states and kinetic pathways, but the qualitative mechanism appears robust.

Figure~\ref{fig:stress_LS}(b) illustrates the dependence of fracture stress on oxide shell thickness. For each oxide thickness, ten independent simulations were conducted to ensure statistical reliability, and the corresponding standard deviations are represented as error bars in the figure. As the oxide layer grows from 0 to 2 nm while consuming the core, the normalized fracture stress decreases from 1.00 for the pristine Si NW to approximately 0.80, 0.77, 0.68, and 0.64 for oxide thicknesses of 0.5, 1, 1.5, and 2 nm, respectively. The initial sharp decline up to 0.5 nm corresponds to the newly formed partially oxidized interfacial region that introduces local bond weakening, as demonstrated in the DFT and MD analyses above. This dramatic decline has also been observed experimentally in native and thermally oxidized Si NWs~\cite{delrio2016near}. \textcolor{msc1}{With continued oxidation, the fracture stress decreases further as the crystalline Si core is progressively consumed, reducing the effective load-bearing cross-section and increasing the tensile stress carried by the remaining core. This trend is therefore expected in part from basic mechanical considerations, including the decreasing crystalline Si core area and the lower stiffness of the amorphous oxide shell. However, the atomistic origin of the degradation is not purely geometric. The simulations show that oxidation simultaneously creates partially oxidized interfacial Si environments that weaken neighboring Si-Si bonds and promote localized stress concentration at the Si-SiO\textsubscript{2} interface. Thus, the mechanical weakening arises from a combined geometric and chemical mechanism: continued oxidation reduces the load-bearing Si core while also creating mechanically vulnerable interfacial bonding environments that facilitate fracture initiation.} This degradation trend is consistent with experimental reports of transient oxidation-induced mechanical softening during the early stages of oxide growth~\cite{grutzik2015oxide}. \textcolor{msc1}{Additional mechanical metrics, including fracture strain and effective Young’s modulus as a function of oxide thickness, are provided in Table S1 of the Supporting Information.} \textcolor{msc1}{Given the underprediction of the Si-O-Si angle by the ReaxFF potential, the absolute magnitudes of oxide-shell stress, stress redistribution, and fracture stress should be interpreted within the quantitative limitations of the force field. However, the qualitative fracture-initiation mechanism remains supported by the observed interfacial stress localization, the reduced Si–Si bond order near partially oxidized interfacial Si environments, and the complementary DFT bond-order calculations showing that oxygen incorporation weakens neighboring Si–Si bonds.}
\begin{figure}[t!]
    \begin{center}
        \begin{tabular}{c}
            \includegraphics[width = 0.8\textwidth]{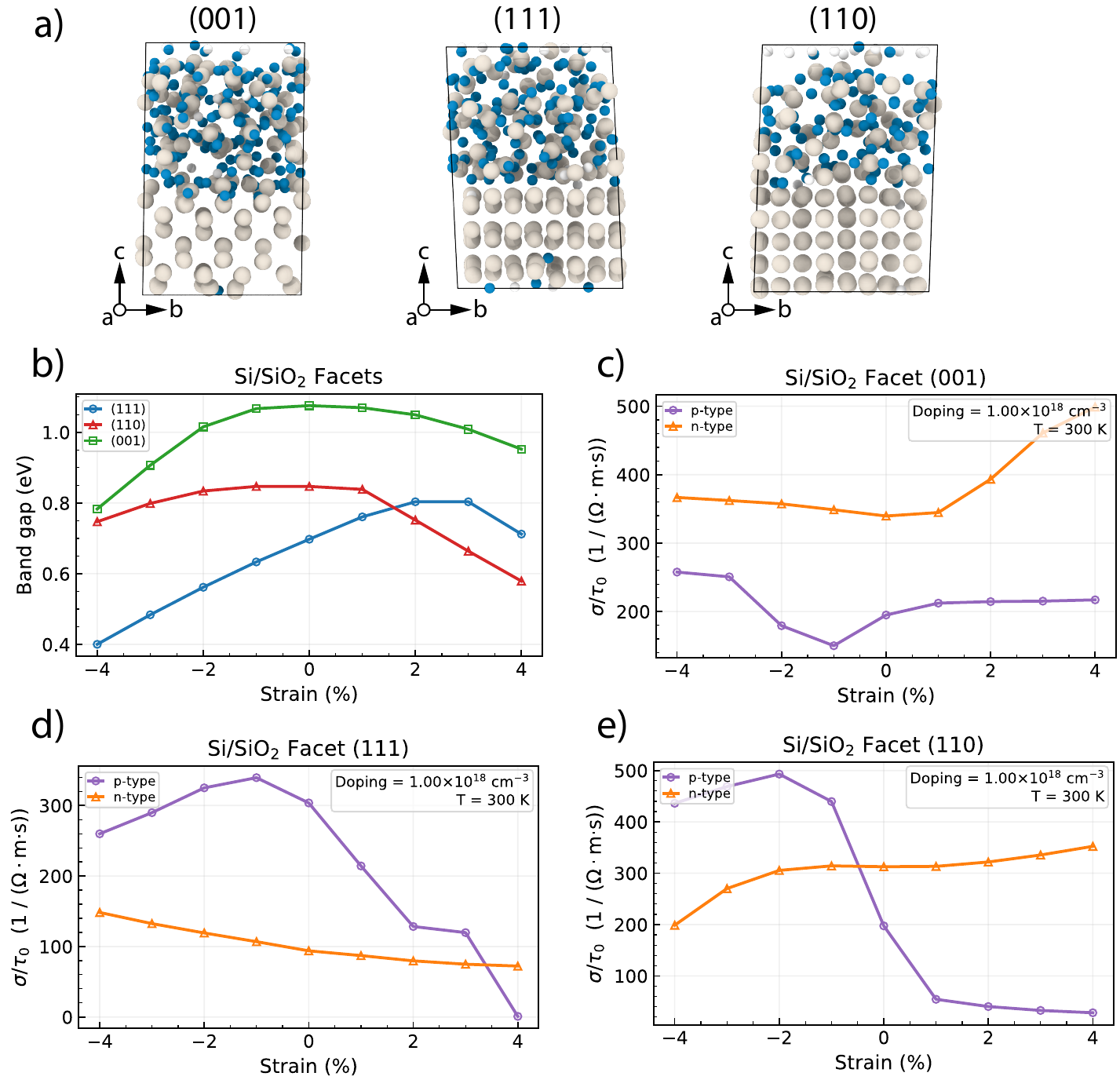}
        \end{tabular}
    \end{center}
    \caption{\label{fig:figdft} \textcolor{msc1}{(a) The Si-SiO\textsubscript{2} crystal structures used in the DFT simulations for the (001), (111) and (110) interfaces. (b) The DFT computed band gap as a function of applied strain for all three facets. DFT and DFT+Boltzmann transport calculation results: The electronic conductivity (with respect to constant relaxation time, \textcolor{msc2}{$\tau_0 = 10^{-14}~s$}) for the c) (001), d) (111) and e) (110) facets for n-type doping and p-type doping as a function of applied strain.}}
\end{figure}
\par

While the rupture analysis establishes the mechanical limitations imposed by the Si-SiO\textsubscript{2} interface, device performance also depends critically on its electronic response under strain. To assess whether the partially oxidized, coordination-disrupted Si environments that govern fracture also contribute to \textcolor{msc1}{facet-dependent strain response of electronic properties}, we turn to DFT-based characterization of facet-specific Si-SiO\textsubscript{2} interfaces. \textcolor{msc1}{These DFT calculations serve as first-principles probes of the electronic consequences of oxidation generated interfacial bonding motifs associated with the mechanically vulnerable Si-SiO\textsubscript{2} regions identified in the reactive MD simulations. Since the MD simulations probe deformation up to fracture whereas the DFT calculations probe moderate static strain states, the two approaches are not directly strain matched, instead they are connected mechanistically through the oxidation-generated interfacial bonding environments. In this approach, facet-resolved periodic planar interface models generated using the same thermal oxidation workflow described in the Methods section are used to represent the local bonding environments characteristic of the crystallographic facets.} \textcolor{msc1}{We consider three planar interfaces oriented along distinct crystallographic planes, (001), (111), and (110). Among these, the (001) and (111) models correspond to the \{100\} and \{111\} surface families exposed by the representative MD NW, whereas the (110) model is included as an additional low-index interface for comparison.} This reduced the atomic complexity and number of atoms in the periodic supercell, enabling DFT-based evaluation of interface properties under applied strain. To alleviate the issue of dangling bonds in amorphous SiO$_2$ inducing mid-gap states (defect-mediated), we passivated the dangling Si bonds with H. We applied strain in the b-direction for the \textcolor{msc1}{(001)}, (111) and (110) interfaces (Fig. \ref{fig:figdft}a), relaxing the atoms at each value of strain from -4 $\%$ to +4 $\%$ strain at increments of 1 $\%$. With regards to the electronic structure, we see \textcolor{msc1}{a directional strain-dependent (positive or negative strain)} response in the band gap of the Si-SiO$_2$ systems for the \textcolor{msc1}{(001), (111) and (110) structures (Fig. \ref{fig:figdft}b).} \textcolor{msc1}{For the (001) interface, there is a steady decrease in band gap for both positive and negative strain.} For the (111) interface, the band gap decreases linearly with compressive strain and increases with small amounts of tensile strain (up to 3 $\%$) and then a slight decrease with higher amounts of tensile strain. For the (110) interface, we observe a sharp decrease in band gap with positive strain and a slight increase (from 0 $\%$ to -1 $\%$) followed by a steady decline (from -1 $\%$ to -4 $\%$) with compressive strain. Specifically, we observe that compressive strain for the (111) interface has a more significant impact on band gap tunability than the (110) case, while the trend is reversed for tensile strain. 

Using our DFT eigenvalues of the strained structures, we performed Boltzmann transport calculations to qualitatively estimate the electronic conductivity ($\sigma$). This $\sigma$ is computed with respect to a constant relaxation time \textcolor{msc2}{($\tau_0 = 10^{-14}~s$)}. These \textcolor{msc1}{$\sigma/\tau_0$} results are depicted in Figs.~\ref{fig:figdft}(c), (d) and (e) for n-type doping and p-type doping (at moderate/heavy electron/hole concentrations of 1.00 $\times$ 10$^{18}$ cm$^{-3}$). \textcolor{msc1}{Unsurprisingly, we see significantly different transport mechanisms dependent on facet orientation and applied strain direction (tensile or compressive). Due to the fact that the constant $\tau_0$ is different for each facet, it is not possible to make a direct quantitative comparison between each orientation, which is why we plot the transport for each facet separately (see Fig. \ref{fig:figdft}(c)-(e)).}

\textcolor{msc1}{Regarding n-type conductivity, we see a slight increase in $\sigma/\tau_0$ with compressive strain for the (001) facet (Fig. \ref{fig:figdft}(c)), while we see a sharper increase for tensile strain (+1~\% to +4~\%).} \textcolor{msc1}{In addition, we observe that the (110) interface (Fig. \ref{fig:figdft}(e))} exhibits an increase in $\sigma/\tau_0$ with tensile strain and a small increase with -1 $\%$ compressive strain, but from -1~\% to -4~\% there is a sharper decrease in electronic conductivity. The response of the (110) interface qualitatively resembles the simulated band gap and conductance of \{110\} Si (no oxide layer) NWs from Ref. \cite{shiri2008strain}. On the other hand, the (111) interface is less responsive to applied strain  for n-type doping, showing a slight decrease with tensile strain and an increase with compressive strain. \textcolor{msc1}{We also observe that hole conductivity is more tunable by strain for the (110) interface than the (111) interface, but both (110) and (111) exhibit similar behavior. In particular, we see opposite behavior in the strain response for hole transport and electron transport in the (110) interface. Contrary to the (110) and (111) facets, we observe that hole conductivity is much less tunable for the (001) facet.}

To support our Boltzmann transport calculations based on DFT, we computed the partial density of states (PDOS) for the \textcolor{msc1}{(001), (111), and (110) under strain (see Fig. S3, Fig. S4, and Fig. S5).} Specifically, we depict the PDOS for 0~\% and for maximum tensile and compressive strain, in addition to strain values where \textcolor{msc1}{we see conductivity maxima or minima (i.e., -2~\% for hole doped (111) or -1~\% for electron doped (001)).} These results are consistent with our conductivity predictions, where a sharp, steeply rising peak near the valence band maximum (VBM) or conduction band minimum (CBM) can correspond to flat bands (low dispersion) and large effective mass/low mobility. In contrast, a more broad, gradually rising peak at the VBM or CBM can correspond to dispersive bands (large curvature) and low effective mass/high mobility. The results for band gap and conductivity demonstrate how the anisotropic bond behavior and mechanical properties (predicted with classical MD) are directly related to \textcolor{msc1}{facet-dependent strain response of electronic properties.} In addition, insights from these transport calculations can provide key insights in designing NW facets with targeted properties modulated by strain.

\section{Conclusion}
In summary, we employed multiscale simulations combining ReaxFF MD and DFT calculations to highlight how oxidation-driven interfacial structure influences both the mechanical response and strain-dependent electronic behavior of Si NWs. Our simulations show that fracture consistently initiates at the Si-SiO\textsubscript{2} interface, where partially oxidized Si bonding environments weaken load-bearing Si-Si networks and promote stress localization within the adjacent crystalline core, underscoring the dominant role of interfacial chemistry and geometry in determining mechanical reliability. Beyond structural failure, the same interface controls \textcolor{msc1}{facet-dependent strain response of electronic properties}: \textcolor{msc1}{DFT calculations reveal that strain-induced modulation of band edges and electronic conductivity ($\sigma/\tau_0$) is remarkably different between (001), (111) and (110) facets, establishing a link between interfacial orientation and strain-driven tunability.} Although the mechanical and electronic analyses were examined independently, both converge on a unified picture in which the Si-SiO\textsubscript{2} interface defines the limits of reliability and performance in strained, oxidized Si NWs. These insights provide practical routes for interface optimization, where minimizing suboxide formation, limiting thermal oxidation and selecting favorable crystallographic facets could enhance both the mechanical robustness and electronic reliability of oxidized Si NW based device architectures.

\section*{Associated Content}
The data generated for this paper are available to the public via
an archive at the following DOI: 10.6084/m9.figshare.34192935.

\subsection*{Supporting Information}
Si NW stress-strain response dependence on strain rates during axial tensile loading. Stress-strain response of independent replicas of Si NW for statistical reproducibility of fracture behaviors. Additional mechanical properties as a function of oxide thickness. Supercell construction and computational details for DFT/DDEC6 bond-order analysis.  DFT model construction details, projected density of states, and orbital contributions for strained Si-SiO\textsubscript{2} interface models.

\section*{Acknowledgment}
All authors thank the National Institute of Standards and Technology for funding, computational, and data-management resources. This work was performed with funding from the CHIPS Metrology Program, part of CHIPS for America, National Institute of Standards and Technology, U.S. Department of Commerce. Certain commercial equipment, instruments, software, or materials are identified in this paper in order to specify the experimental procedure adequately. Such identifications are not intended to imply recommendation or endorsement by NIST, nor it is intended to imply that the materials or equipment identified are necessarily the best available for the purpose.

\bibliography{3_references}
\end{document}


\setcounter{page}{1}
\renewcommand{\thepage}{S\arabic{page}}

\renewcommand{\figurename}{Figure} 
\renewcommand{\thefigure}{S\arabic{figure}}

\renewcommand{\tablename}{Table}
\renewcommand{\thetable}{S\arabic{table}}

\section*{Effect of Strain Rate on Stress-Strain Response}
\begin{figure}[t!]
    \begin{center}
        \begin{tabular}{c}
            \includegraphics[width = 0.8\textwidth]{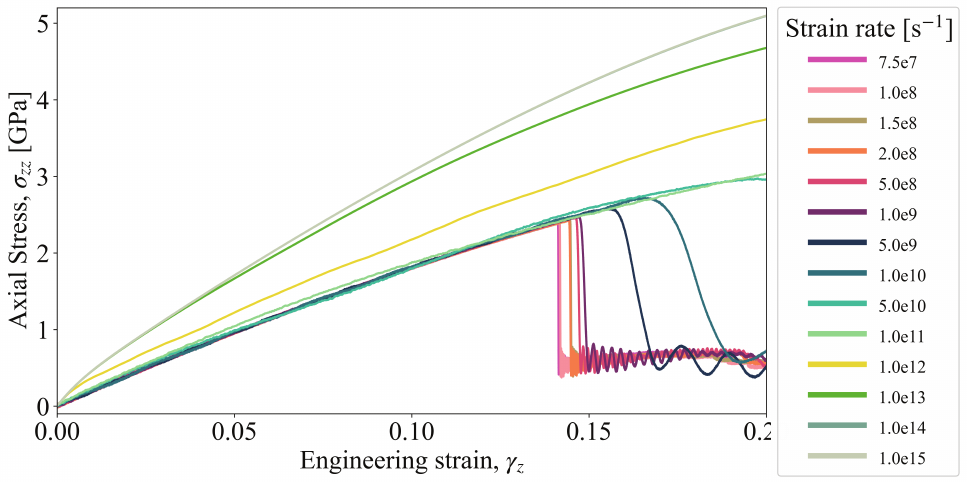}
        \end{tabular}
    \end{center}
    \caption{\label{fig:SSvsSR}Stress Strain response for a Si nanowire (Si NW) with 2 nm thick oxide shell for different strain rates.}
\end{figure}

To assess the sensitivity of the tensile response to strain rate, we performed uniaxial loading simulations of oxidized Silicon nanowires (Si NWs), with oxide thickness 2~nm across fourteen different strain rates, ranging from $7.5 \times 10^{7}$ s$^{-1}$ to $1 \times 10^{15}$ s$^{-1}$. The resulting stress–strain curves are shown in Fig.~\ref{fig:SSvsSR}. At the lowest rates ($\leq 2 \times 10^{8}$ s$^{-1}$), the wires exhibit a distinct brittle-like rupture, characterized by an abrupt stress drop following elastic loading. This abrupt rupture mechanism is consistent with fracture events localized at the Si–SiO\textsubscript{2} interface, as discussed in the main text.  

At higher strain rates, the fracture behavior changes qualitatively. Instead of sudden rupture, the NWs undergo progressive stretching and void nucleation, with stress relaxation occurring gradually rather than catastrophically. This response reflects the inability of the system to fully relax atomic configurations when strained too rapidly, thereby altering the apparent failure mode.  

Based on these observations, we selected a strain rate of $1 \times 10^{8}$ s$^{-1}$ for all subsequent simulations presented in the main manuscript. This rate is sufficiently slow to capture abrupt fracture events consistent with interfacial rupture, while remaining computationally tractable.
\begin{figure}[t]
    \begin{center}
        \begin{tabular}{c}
            \includegraphics[width = 0.6\textwidth]{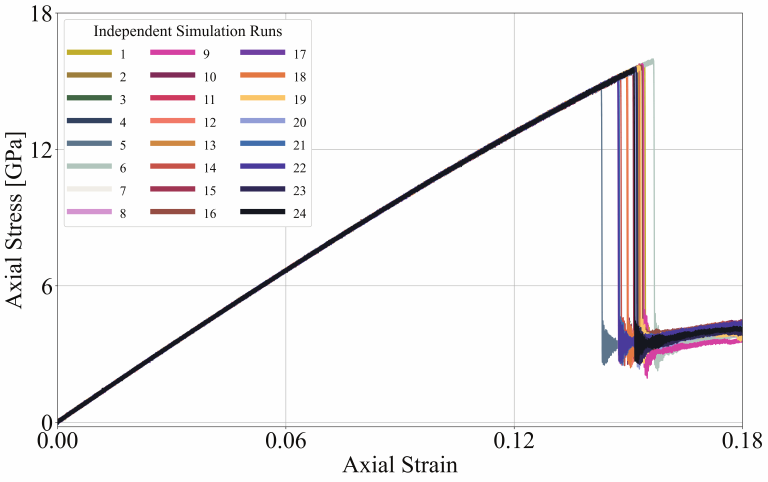}
        \end{tabular}
    \end{center}
    \caption{\label{fig:SSreplicas}Stress Strain response of independent simulation runs for a Si NW with 2~nm thick oxide shell for a strain rate of $10^{8}$ s$^{-1}$.}
\end{figure}
\section*{Independent Simulation Runs}
To assess the statistical reproducibility of fracture initiation and propagation, we conducted 24 independent uniaxial tensile simulations on oxidized Si NWs using distinct initial velocity seeds. Figure~\ref{fig:SSreplicas} presents the corresponding stress--strain responses for all replicas. Across these simulations, rupture consistently initiated at the Si-SiO\textsubscript{2} interface, either along a single facet or at the junction between two neighboring facets, confirming that interfacial failure is not a stochastic artifact of a single trajectory. \textcolor{msc}{A classification of the rupture initiation sites showed that 14 of the 24 replicas initiated at or near facet junctions, while 10 initiated along individual facets. Therefore, rupture was consistently interfacial but did not occur exclusively at a single crystallographic junction or at only one class of interfacial site.} The mean fracture stress was 15.51$\pm$0.25~GPa, and the corresponding fracture strain was 15.05$\pm$0.3~\%. Minor variations in the pre-yield region reflect thermal fluctuations in atomic stress and slight differences in oxidation topology, but the overall deformation and rupture mechanisms remained identical across all replicas. \textcolor{msc}{The nonzero stress observed after the sharp stress drop is indicative of how the surrounding oxide shell is still partially connected and continues to carry a reduced tensile load during further deformation. During this post-rupture stage, the oxide shell undergoes progressive stretching, void nucleation, and eventual separation. This response is consistent with deformation in amorphous SiO\textsubscript{2}, where failure can proceed through gradual void formation and network rearrangement rather than through a single atomically sharp cleavage event~\cite{chowdhury2016molecular}. These results demonstrate that the interface-localized failure mode observed in the main text is statistically robust and independent of initial dynamic perturbations.}
\begin{table}[h]
\captionsetup{labelfont={color=msc},textfont={color=msc}}
\color{msc}
\centering
\caption{\label{tab:amorphousSiO2comparison} Mechanical properties of oxidized Si NWs as a function of oxide shell thickness. Fracture stress and fracture strain were obtained from the peak stress and corresponding strain in the tensile stress-strain response, respectively. The effective axial Young’s modulus was calculated from the initial linear elastic region of the stress-strain curve (up to strain 2\%). These values quantify the influence of oxidation on both the elastic response and failure behavior of the NW.}
\begin{tabular}{cccc}
\hline\hline
\shortstack{Oxide\\thickness (nm)} & \shortstack{Fracture\\stress (GPa)} & \shortstack{Fracture\\strain} & \shortstack{Effective Young's\\modulus (GPa)}\\
\hline
0   & 24.32 $\pm$ 0.42 & 0.147 $\pm$ 0.003 & 200.74 $\pm$ 0.06 \\
0.5 & 19.35 $\pm$ 0.29 & 0.142 $\pm$ 0.003 & 161.47 $\pm$ 0.27 \\
1.0 & 18.75 $\pm$ 0.30 & 0.146 $\pm$ 0.003 & 147.05 $\pm$ 0.18 \\
1.5 & 16.48 $\pm$ 0.15 & 0.143 $\pm$ 0.001 & 128.18 $\pm$ 0.11 \\
2.0 & 15.63 $\pm$ 0.11 & 0.152 $\pm$ 0.002 & 114.13 $\pm$ 0.12 \\
\hline\hline
\end{tabular}
\end{table}

\newpage

\section*{Density Functional Theory}
\begin{figure}[th!]
    \begin{center}
        \begin{tabular}{c}
            \includegraphics[width = 0.99\textwidth]{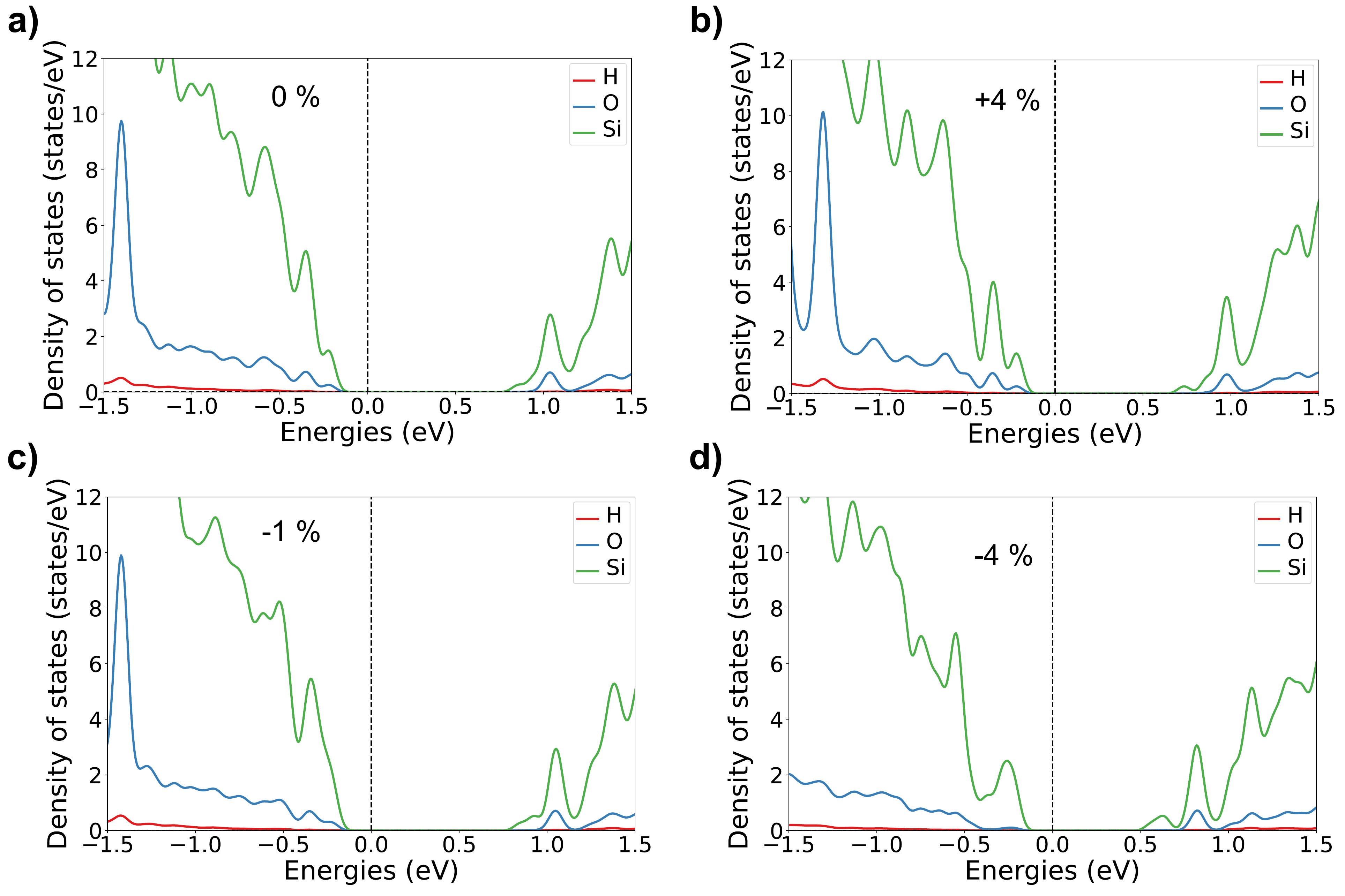}
        \end{tabular}
    \end{center}
    \caption{\label{fig:si-dft1}} \textcolor{msc}{The partial density of states for the Si-SiO$_2$ (001) interface under various amounts of strain: a) 0~\%, b) +4~\%, c) -1~\%, and d) -4~\%.}
\end{figure}

\begin{figure}[th!]
    \begin{center}
        \begin{tabular}{c}
            \includegraphics[width = 0.99\textwidth]{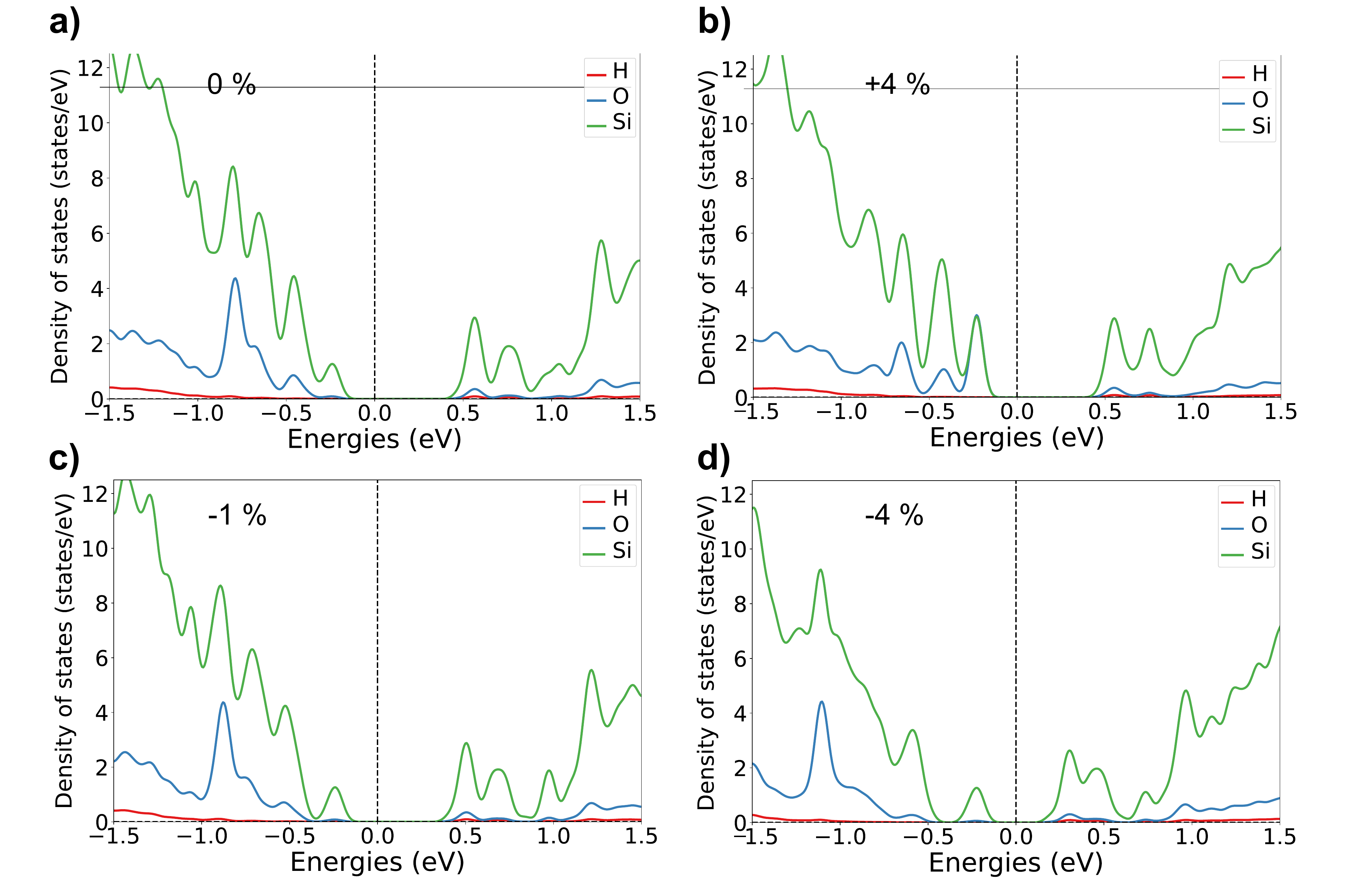}
        \end{tabular}
    \end{center}
    \caption{\label{fig:si-dft2}} The partial density of states for the Si-SiO$_2$ (111) interface under various amounts of strain: a) 0~\%, b) +4~\%, c) -1~\%, and d) -4~\%.
\end{figure}

\begin{figure}[th!]
    \begin{center}
        \begin{tabular}{c}
            \includegraphics[width = 0.99\textwidth]{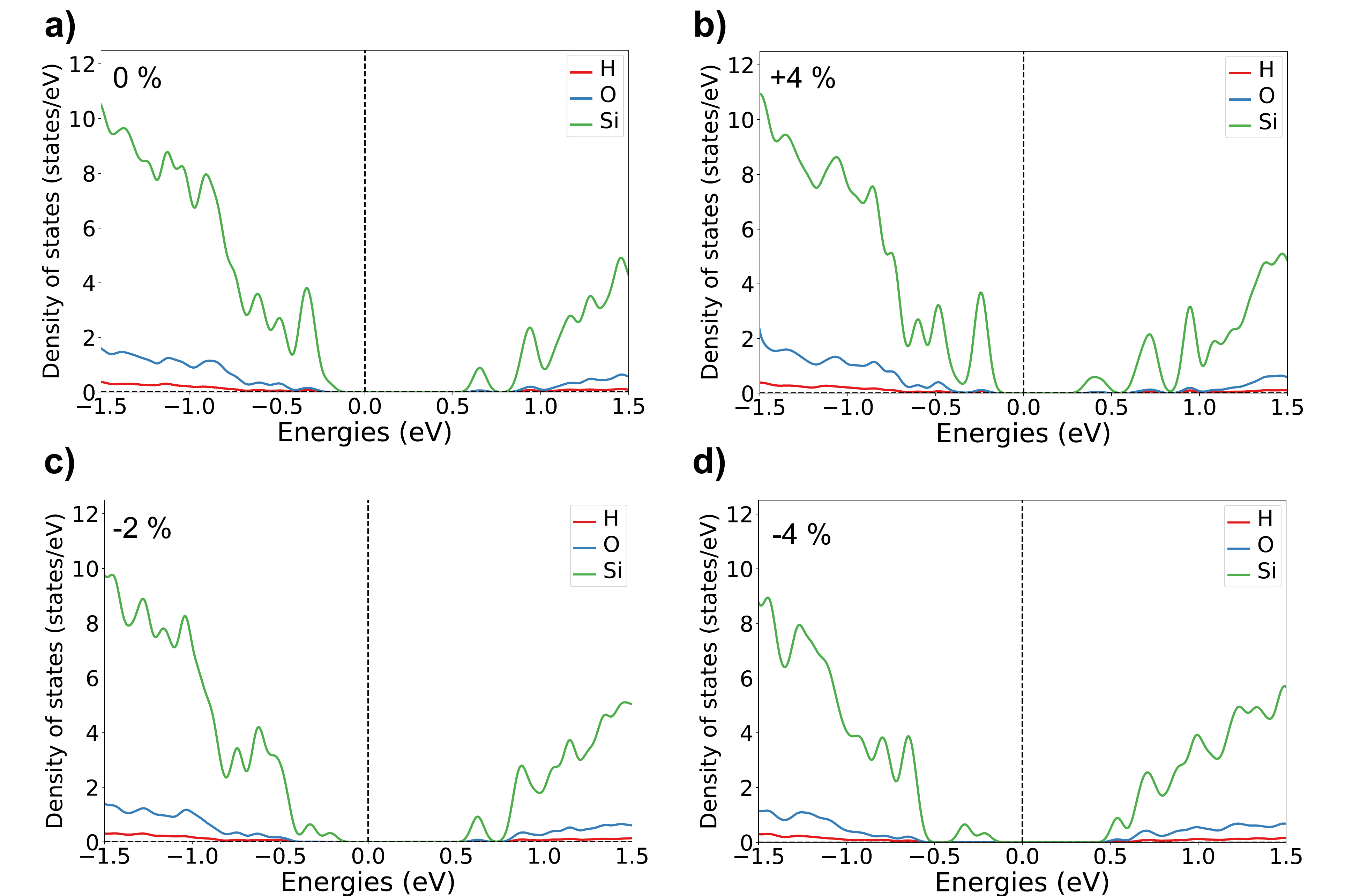}
        \end{tabular}
    \end{center}
    \caption{\label{fig:si-dft3}} The partial density of states for the Si-SiO$_2$ (110) interface under various amounts of strain: a) 0~\%, b) +4~\%, c) -2~\%, and d) -4~\%.
\end{figure}

In this work, we performed two types of density functional theory (DFT) calculations:

\begin{enumerate}
\item \textbf{Si-SiO$_2$ interfaces (NW facets).} We carried out DFT calculations for Si-SiO$_2$ interfaces with multiple orientations representative of the nanowire facets. Computational details and results are provided in the main text (Fig. 5), and supplemental partial densities of states are shown in Fig. \ref{fig:si-dft1} - \ref{fig:si-dft3}.

\item \textbf{Crystalline supercells with interstitial O (local oxidation probe).} We computed bond lengths and bond orders in crystalline Si supercells containing interstitial O to emulate local interfacial oxidation. Calculations were performed with the Vienna \textit{Ab initio} Simulation Package (VASP) \cite{PhysRevB.54.11169} using the projector augmented-wave (PAW) method \cite{PhysRevB.59.1758} and the PBE exchange–correlation functional \cite{perdew1996generalized}. A 2 x 2 x 2 conventional Si supercell (64 atoms) with up to three interstitial O atoms was employed. Brillouin-zone sampling used a $\Gamma$-centered Monkhorst-Pack 10 x 10 x 10 grid and a plane wave kinetic-energy cutoff of 600 eV \cite{1976-kpoints}. Electronic self-consistency was converged to 10$^{-6}$ eV, and both lattice vectors and internal coordinates were fully relaxed. Bond orders were obtained \textit{a posteriori} from the converged charge density using the Chargemol code (DDEC6 method), which partitions the total electron density into chemically meaningful atoms-in-materials and computes bond orders directly from the electron density in a robust, basis-set–independent real-space framework \cite{chargemol}. For each configuration (Si, Si\textsuperscript{1+}, Si\textsuperscript{2+}, and Si\textsuperscript{3+}) we report the average Si-Si and Si-O distances and the corresponding DDEC6 bond orders (Table 2).
  
\end{enumerate}

\clearpage
\bibliography{3_references}